\documentclass[twocolumn,showpacs,preprintnumbers,amsmath,amssymb,prb,aps]{revtex4-1}
\usepackage{epsfig}% Include figure files$
\usepackage{epstopdf}
\usepackage{graphicx}% Include figure files
\usepackage{dcolumn}% Align table columns on decimal point
\usepackage{bm}% bold math
\usepackage{textcomp}
\usepackage{array}
\usepackage{caption}
\usepackage{xfrac}
\usepackage{mathtools}
\usepackage{amsmath}
\usepackage{subfigure}

\begin{document}

\title{An \textit{ab-initio} study of topological and transport properties of YAuPb}
\author{Vivek Pandey$^{1}$ }
\altaffiliation{vivek6422763@gmail.com}
\author{Antik Sihi$^{1}$ }
\author{Sudhir K. Pandey$^{2}$}
\altaffiliation{sudhir@iitmandi.ac.in}
\affiliation{$^{1}$School of Basic Sciences, Indian Institute of Technology Mandi, Kamand - 175075, India\\
$^{2}$School of Engineering, Indian Institute of Technology Mandi, Kamand - 175075, India}
\date{\today}

\begin{abstract}

     In the last few decades, the study of topological materials has been carried out on an extensive scale. Half-Heusler alloys are well known for their topological behaviours. In this work, we present a detailed study of topological properties of a ternary Half-Heusler alloy, YAuPb, using the tight-binding approach. We have calculated some important topological properties which includes$-$ finding nodes and their chiralities, Berry curvature ($\boldsymbol\Omega$) and the surface-states. 5 pairs of characteristic nodes with equal and opposite chiralities are obtained. Based on the study of these properties, we categorise  the material as non-trivial topological semimetal. Besides the topological behaviours, we present a comparative study of temperature dependent transport properties corresponding to the chemical potential ($\mu$) of the Fermi level and the node points. The temperature range chosen for the study is 50-300 K. The results obtained from the calculations of electrical conductivity per unit relaxation time ($\boldsymbol\sigma/\tau$) and the electronic part of thermal conductivity per unit relaxation time ($\boldsymbol\kappa_0$) indicates the conducting nature of the material to both the heat and electricity. At the Fermi level, the value of Seebeck coefficient ($S$) is found to be $\sim -9.07 \hspace{1mm}(-35.95)$ $\mu$V$ K^{-1}$ at 50 (300) K. The negative value of $S$ indicates the n-type behaviour of the compound. The calculated value of electronic specific heat (Pauli magnetic susceptibility) corresponding to Fermi level is $\sim 0.03 \hspace{1mm}(0.18) \times 10^{-2}$ $ Jmol^{-1}K^{-1}$ ($\sim 1.21 \hspace{1mm}(1.14) \times 10^{-10}$ $ m^{3}mol^{-1}$) at 50 (300) K. This work suggests that YAuPb is a promising candidate of non-trivial topological semimetals which can be employed in transmission of heat and electricity, and as n-type material within the temperature range of 50-300 K.

\end{abstract}

\maketitle

\section{Introduction} 
\setlength{\parindent}{3em}
        Topological materials are gathering the attention of researchers from all over the world in current scenario.
  They are found to be extremely useful due to their varying topological and transport behaviours. The branch of condensed matter physics dealing with these materials evolved from the study of topological insulators in its early stage. This field further advanced with the observation and study of other topological materials like topological semimetals and even metals\cite{Burkov}. In the past few decades, it is seen that the study of electronic states derived topology of the materials has been carried out on an extensive scale. It is observed that these materials are associated with different topological invariants which can be calculated using their bulk dispersion curve \cite{Nat,Roy}. The curve also reflects some of the important signatures associated with these materials. For instance, it has been reported in several works that the band inversion in topological materials usually occurs near the $\Gamma$-point of the Brillouin zone (BZ)\cite{Zhang,X,Y}. Such inversion of bands are responsible for the change in the topological order of the materials \cite{FuKane}. The inversion of bands occurs during the band-splitting. This splitting may result in the creation of several types of $\Gamma$-points. These are- s-like $\Gamma_6$ with two-fold degeneracy, p-like $\Gamma_7$ with two-fold degeneracy and p-like $\Gamma_8$ with four-fold degeneracy. The presence of s-like $\Gamma_6$ above the p-like $\Gamma_8$ in the dispersion curve represents the normal ordering of the bands. On the other hand, presence of p-like $\Gamma_8$ above the s-like $\Gamma_6$ represents the inversion in the natural ordering of the bands \cite{Lin}.  The presence of band-inversion in the dispersion curve usually indicates that the compound is non-trivial topological insulator or metal \cite{FuKane1}. 
  \par The observation of non-trivial topological electronic structure in metals have been known from the long time \cite{Haldane,Murakami}. Furthermore, the overlap between the conduction and valence bands in metals is also well-known. Unlike metals, in case of semimetals, the topmost valence band and the bottom-most conduction band touches at only specific values of the crystal momentum in the BZ. These touching points usually lie at the Fermi level and are referred to as nodes. If the points are large in number, located close to each other in a given direction of the BZ, they are collectively referred as nodal line\cite{Burkov}. The existence of node points in the dispersion curve was observed long before their importance in the study of material could be identified. In the recent past, many important properties related to semimetals have been identified and extensively studied. Some of these are chirality of nodes\cite{Zyuzin}, chiral anomaly\cite{Zyuzin}, quantum spin-Hall effect\cite{Sun}, Fermi arc\cite{Morali}, etc. These studies result in the theoretical discovery of Weyl semimetals \cite{8,9,10} and later on the Dirac semimetals \cite{11,12}. At present, these are the two broadly divided class of semimetals. With the experimental realization of these materials \cite{1,2,3}, the field has now emerged to a prominent position in the quantum condensed-matter research.
\par Half-Heusler alloys are one of the common materials which show topological behaviours. These materials are generally represented by the chemical formula XYZ, where X and Y typically belongs to the transition elements and Z belongs to the class of heavy elements\cite{Xiao}. These materials are becoming prominent in current research due to their various crystal structures and different electronic, magnetic, and transport properties. The study of the properties like electronic structure, phonon modes and the prediction of figure of merit, etc. are usually carried out for these alloys\cite{S1}. In addition to this the studies have been also performed to explore the thermoelectric transport in these materials\cite{S2}. The half-Heusler alloys with general formula \textit{R}AuPb (\textit{R}=Gd,Er,Ho,Y,Tb,Dy) possess face-centred cubic crystal structure similar to MgAgAs type\cite{Melnyk}. They belong to the space group \textit{F}-43\textit{m} \cite{Marazza}. Majority of MgAgAs type Half-Heusler alloys are composed of rare-earth elements. These elements provide the additional properties to these alloys. The presence of these elements may make the alloys superconducting or magnetic in nature. This behaviour is responsible for providing the different temperature dependent transport properties in these compounds\cite{Go}.

\par The compound YAuPb is also a candidate of ternary half-Heusler alloys, RAuPb. Several researches have been already performed on this compound. For instance, the structural properties of the material were studied to determine its lattice parameters\cite{Marazza} and the bulk modulus\cite{Kandpal}. Also, a comparative study of the variations of transport properties with temperature was carried out for YAuPb and LuAuPb under 5\% strain. In this work, the properties like Seebeck coefficient, electronic thermal conductivity and electrical conductivity along with the power factor and the figure of merit were studied\cite{Singh}. Furthermore, the band-characters in the vicinity of Fermi level were also studied. It was found that the dispersion curve of the material reflects the band-inversion and hence it was proposed to be topological metal\cite{Lin}. However, the orbitals responsible for the inversion of bands were not explicitly mentioned. Later on, the band-ordering\cite{Lekhal} and the electronic properties\cite{Sawai} of the compound were also studied. In this study, the compound was found to show semi-metallic behaviour. It is important to note that the compound has been already claimed to be topological or semi-metallic in nature, which might be based on the observation of band-inversion, a signature of topological materials. In spite of that, its topological properties are not explored so far in greater details. Moreover, the transport properties of the compound in its pure form are also not much studied. Thus, the need of topological analysis of YAuPb along with the study of its transport properties is very necessary for better understanding of the material and its practical applications. 

\par In the study of the electronic properties of the compound, band inversion is observed near the $\Gamma$-point. The orbitals responsible for the inversion of bands are also obtained. Furthermore, in the study of the topological properties of the material, 5 pairs of characteristic nodes having equal and opposite chiralities are obtained in the vicinity of the Fermi level. These node points are found to be monopoles of the Berry curvature ($\boldsymbol\Omega$). In addition to this, several surface states for the slab system corresponding to the top and bottom surface of the unit cell are also obtained for the material. Apart from the topological behaviours, a comparative study of the temperature dependent transport properties corresponding to the chemical potential ($\mu$) of the node points and Fermi level are also carried out. The temperature range chosen for this calculation is 50-300 K.
 The properties under study includes - electrical conductivity per unit relaxation time ($\boldsymbol\sigma/\tau$), electronic part of thermal conductivity per unit relaxation time ($\boldsymbol{\kappa_0}$), Seebeck coefficient (\textit{S}), electronic specific heat (\textit{c}), Pauli magnetic susceptibility ($\chi$) and Hall coefficient ($R_H$). The calculation shows that the material have conducting nature to both the heat and electricity at the Fermi level ($\mu_F$), within the given temperature range. The calculated value of $\boldsymbol {\sigma}/\tau$ corresponding to $\mu_F$ is found to be $\sim 1.36 \hspace{1mm}(1.54)\times 10^{19}$  $\Omega^{-1}m^{-1}s^{-1}$ at 50 (300) K . Also, the value of $\boldsymbol\kappa_0$ corresponding to $\mu_F$ is found to be  $\sim 0.2 \hspace{1mm}(1.5) \times 10^{14}$ $ Wm^{-1}K^{-1}s^{-1}$ at 50 (300) K. In addition to this the calculated value of \textit{S} corresponding to $\mu_F$ is obtained to be $\sim -9.07 \hspace{1mm}(-35.95)$ $\mu$V$ K^{-1}$ at 50 (300) K. The negative values of $S$ indicates the n-type behaviour of the material at this $\mu_F$.

\begin{figure}
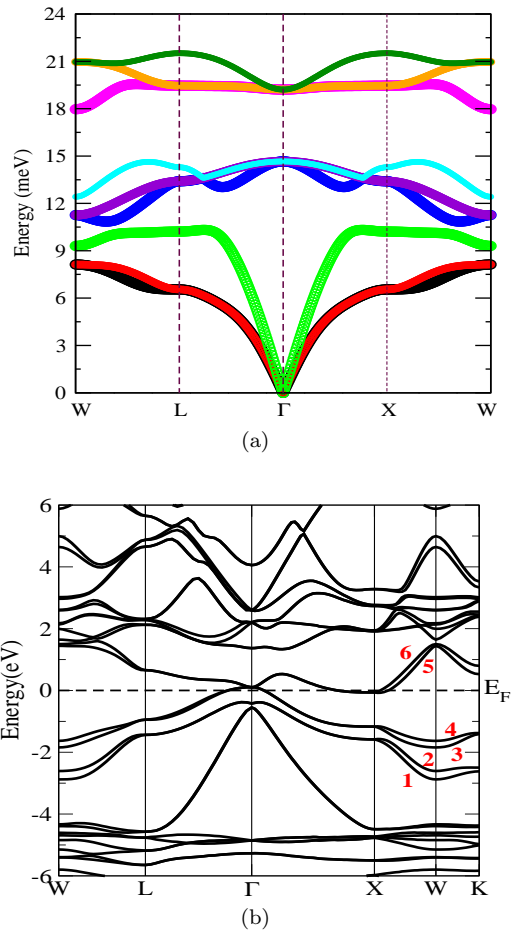

    \centering
    \subfigure[]
    {
        \includegraphics[width=0.75\linewidth, height=5.5cm]{fig1a.eps}
    }
    \subfigure[]
    {
        \includegraphics[width=0.79\linewidth, height=5.7cm]{fig1b.eps}
    }
    \caption
    { { The plots (a) and (b) show the phonon and electronic dispersion curves of YAuPb, respectively.}
    }
    \label{fig:foobar}
\end{figure}

%***************************************** Computational  **************************************

\section{Computational details}

   The density functional theory (DFT) based computations are performed using WIEN2k \cite{WIEN2k} package. The non-magnetic calculations of YAuPb are carried out using full-potential linearized augmented plane wave method. The PBESol, which is based on Generalised Gradient Approximation (GGA), is used as exchange-correlation functional in this calculation\cite{Perdew}. The space group and the lattice constant of the compound are \textit{F}-43\textit{m} and 6.729 \AA, \hspace{1mm}respectively. The Wyckoff positions of the atoms Y, Au and Pb are 4c (1/4,1/4,1/4), 4a (0,0,0) and 4d (3/4,3/4,3/4), respectively \cite{Marazza}. The muffin-tin sphere radii used for each atom is 2.5 Bohr. 10 $\times$ 10 $\times$ 10 k-mesh is used with the energy convergence limit equal to 10$^{-4}$ Ryd. The phonon calculation of this compound is carried out by PHONOPY\cite{Phonopy} code using finite displacement method with making 2 × 2 × 2 supercell. Moreover, 5 × 5 × 5 k-mesh is used for the force calculation using WIEN2k code, where the force convergence criteria is fixed at 0.1
mRy/Bohr.\par The Maximally localized Wannier functions (MLWF) based tight-binding (TB) model is obtained using WANNIER90 \cite{Wannier90}. This code is interfaced with the WIEN2k package. To have a good accuracy in the study of the compound, the model is obtained on k-mesh of 15 $\times$ 15 $\times$ 15 size. This model is further used to study the topological properties of the compound using WannierTools code\cite{WannierTools}. The calculations of transport properties are performed using the BoltzTraP code\cite{BoltzTraP}. These properties are highly sensitive to the behaviour of the band structure. In order to maintain a good accuracy in determining the properties, the calculations of ground-state eigenvalues are done on a dense k-mesh of 50 × 50 × 50 size. The calculations of eigenvalues are performed using self-consistent method with the charge convergence limit/cell equals to 10$^{-4}$ electronic charges.

\section{Results and Discussion} 
\subsection{Electronic structure and topological properties}
\setlength{\parindent}{3em}
\setlength{\parskip}{0.2em}

The structural stability of any compound consisting the heavy elements in cubic phase is normally questionable. Here, it is noted that this compound is already experimentally synthesized in cubic phase. However, further to check the mechanical stability of this compound, the phonon band structure of YAuPb along the high symmetric directions \textit{W-L-$\Gamma$-X-W} are shown in Fig. 1(a). At this point, it is interesting to note that no negative frequency is observed form this phonon band structure plot, which indicates the presence of mechanical stability of this compound in cubic phase. Moreover, the first two acoustic phonon branches are degenerate along \textit{L-$\Gamma$-X} direction. Third acoustic branch meets with the other two acoustic phonon branches around the $\Gamma$-point. Moreover, six optical phonon branches are seen from this figure which are not degenerate with the acoustic branches. Moving further, the plot of the electronic dispersion of YAuPb is shown in Fig. 1(b). The compound is composed of heavier elements like Au and Pb. As the effect of SOC is considerably high in heavier elements, hence it is included in obtaining the dispersion curve. The high symmetric directions along which the curve is obtained are \textit{W-L-\textit{$\Gamma$}-X-W-K} of the BZ. It is seen that the bands are highly entangled in a wide energy range, from -6.0 eV to 6.0 eV. The different transport properties and topological behaviours of any compound are highly sensitive to the states close to Fermi level. Hence, bands marked from 1-6, in the figure, play an important role in deciding the characteristics and behaviours of the compound under study. The bands 1 and 2, which are non-degenerate along \textit{W-L} direction, become degenerate along \textit{L-\textit{$\Gamma$}-X} direction. The degeneracy of these bands again gets lifted up along \textit{X-W-K} direction. Similarly, bands 3 and 4, which are non-degenerate along \textit{W-L} direction, become degenerate along \textit{L-\textit{$\Gamma$}-X} direction. These bands get splitted-up along \textit{X-W-K} direction. These bands gradually come closer to one-another along \textit{W-K} direction. They become degenerate at the \textit{K}-point in the BZ. Also, as can be seen form Fig. 1(b), the bands cross the Fermi level in the vicinity of \textit{$\Gamma$}-point. Furthermore, the bands 5 and 6, which are degenerate at \textit{W}-point, become non-degenerate along \textit{W-L} direction. At \textit{L}-point, the bands again become degenerate. These bands continue to be degenerate along \textit{L-\textit{$\Gamma$}-X} direction. At \textit{X}-point, the bands get splitted-up. Moreover, these bands remain non-degenerate along \textit{X-W} direction. At \textit{W}-point, the bands 5 and 6 again become degenerate. In addition to this, along \textit{W-K} direction the bands remain non-degenerate. It is seen from the figure, that the bands 3 to 6 are degenerate in the vicinity of the \textit{$\Gamma$}-point, at $\sim$100 meV above the Fermi level. Moreover, due to the presence of partially filled 4d electrons in Y atom, one may expect the importance of on-site \textit{U} value in studying the electronic structure of this compound. In light of this, considering two different values of \textit{U} parameters, the DFT+\textit{U} calculations are also performed for YAuPb with fixing \textit{U}=1 eV and 2 eV, respectively. After these calculations, the usual shifting of the bands towards the higher energy is observed in the band structure of DFT+\textit{U} as compare to DFT’s band structure. The value of this shifting becomes more with increasing in the value of \textit{U} parameter. However, it is difficult to find the appropriate value of \textit{U} to describe the electronic structure of YAuPb. Recent theoretical work shows that the value of fully screened Coulomb interaction (\textit{W}) is preferred to use as \textit{U} parameter in DFT+\textit{U} calculation \cite{Sihi}. Thus in this direction, more systematic studies are needed for this sample in future. Therefore, to avoid the confusion of parameters, the DFT method is chosen for the present study. In general, the bulk dispersion curve directly depicts the important behaviour of topological properties by studying the band inversion of a given compound. Hence, a further detailed study of the dispersion curve is discussed below.

\begin{figure}
    \centering
    \subfigure[]
    {
        \includegraphics[width=0.45\linewidth, height=4.5cm]{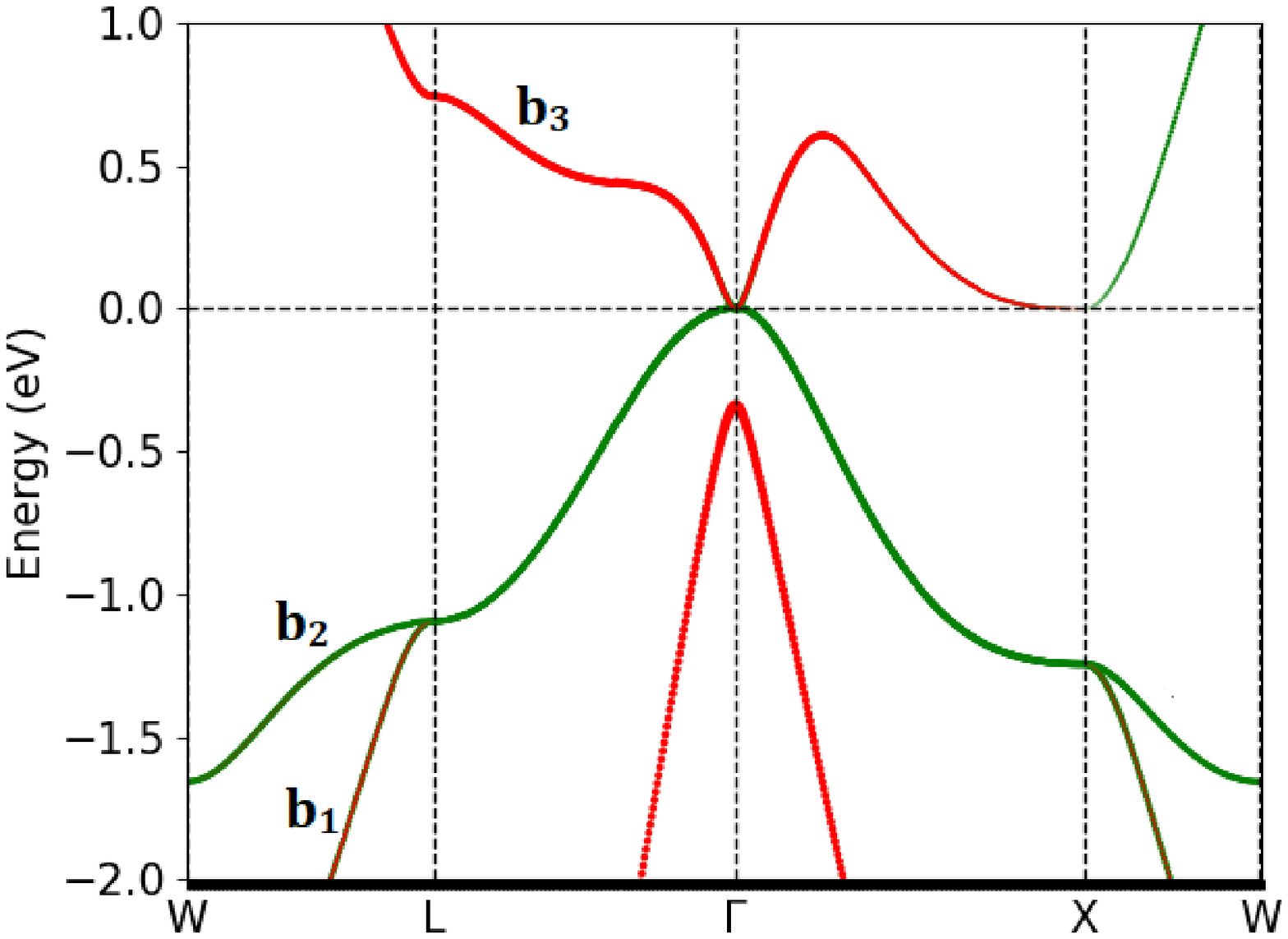}
    }
    \subfigure[]
    {
        \includegraphics[width=0.45\linewidth, height=4.5cm]{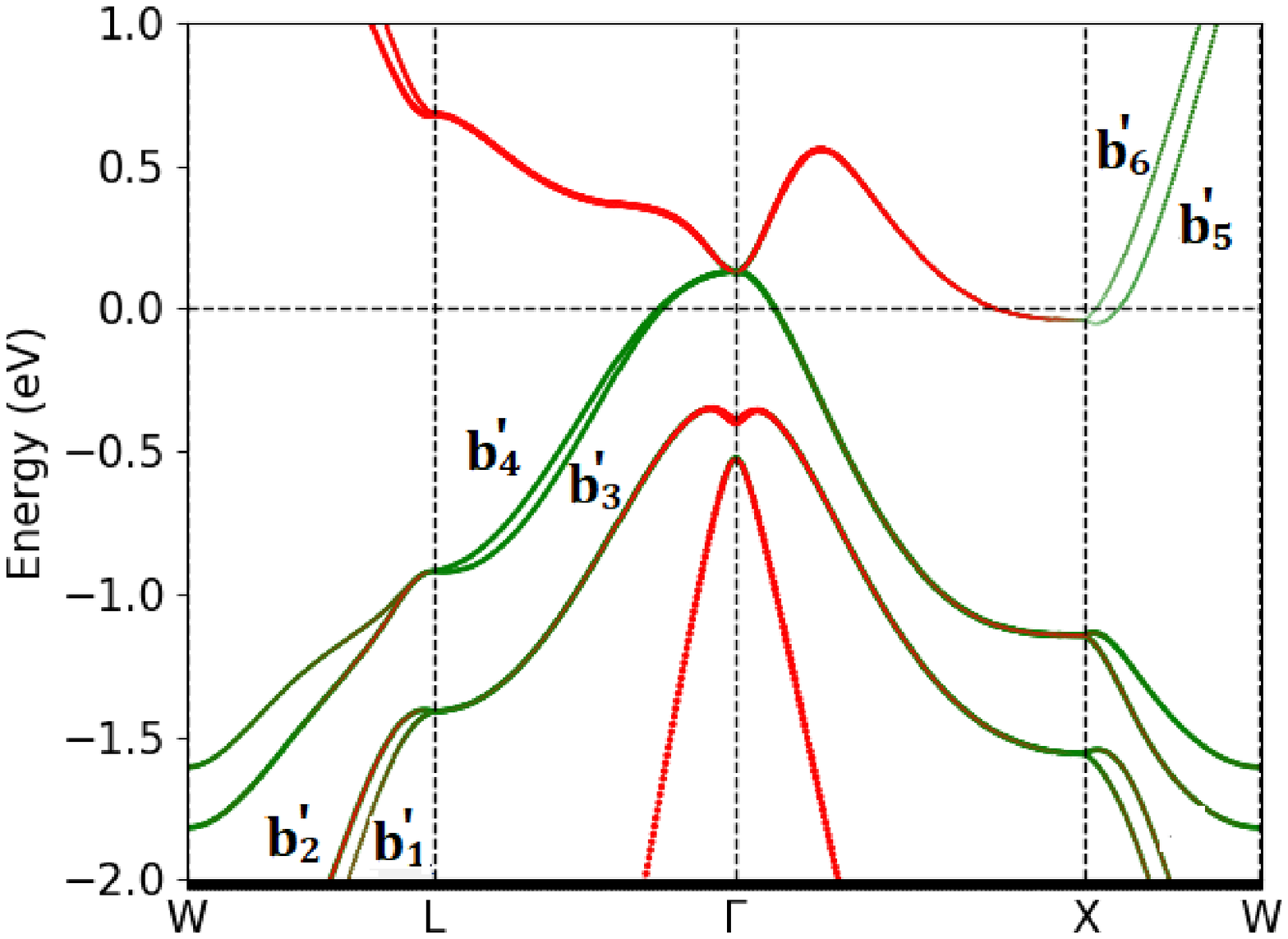}
    }
    \caption
    { {\footnotesize The plot (a) and (b) shows the dispersion curve without SOC and with SOC, respectively. The red (green) colour denotes the contribution of \textquoteleft5\textit{p}' character of Au (\textquoteleft6\textit{s}' character of Pb).}
    }
    \label{fig:foobar}
\end{figure}

\par The presence of band inversion, in the dispersion curve is a signature of the compounds showing non-trivial topological behaviours \cite{Lin}. With this motivation, the band character near the Fermi level and the effect of SOC on it are studied for the compound. In light of this, the dispersion curves with SOC and without SOC are compared in Fig. 2. The intensity of red and green colour denotes the contribution of  \textquoteleft5\textit{p}' character of Au and the \textquoteleft6\textit{s}' character of Pb, respectively.
Fig. 2(a) shows the dispersion curve in the absence of SOC. It is observed that the bands $b_{1}$, $b_{2}$ and $b_{3}$ are degenerate at the \textit{$\Gamma$}-point. In the vicinity of \textit{$\Gamma$}-point, the band $b_{2}$, which has the major contribution from \textquoteleft5\textit{p}' orbitals of Au, is below the band $b_{3}$, whose major contribution comes from \textquoteleft6\textit{s}' orbitals of Pb. This follows the natural order of \textit{s}- and \textit{p}-type-orbital-derived band
structure. After the addition of SOC, the bands $b_{1}$, $b_{2}$ and $b_{3}$ get splitted-up into $b_{1}^{'}$ \& $b_{2}^{'}$, $b_{3}^{'}$ \& $b_{4}^{'}$, and $b_{5}^{'}$ \& $b_{6}^{'}$, respectively. These bands are shown in Fig. 2(b). It is seen that the bands $b_{1}^{'}$ and $b_{2}^{'}$ are two-fold degenerate at the \textit{$\Gamma$}-point, below the Fermi level. The bands $b_{3}^{'}$ to $b_{6}^{'}$  are four-fold degenerate at the \textit{$\Gamma$}-point, $\sim$100 meV above the Fermi level. On close observation of the characters in the vicinity of the \textit{$\Gamma$}-point, it is seen that the contribution to the four-fold degenerate state comes from the \textquoteleft5\textit{p}' orbitals of Au. On the other hand, the contribution to the two-fold degenerate state comes from the \textquoteleft6\textit{s}' orbitals of Pb. This behaviour represents an inversion to the natural order of \textit{s}- and \textit{p}-type-orbital-derived band structure. Also, the presence of band inversion in the compound gives a strong reason to study its topological properties.

\par TB method is one of the widely used approach to study the topological properties of any compound due to its less computational cost. It is already mentioned that most of the topological properties are sensitive to the states near the Fermi level. Therefore, a TB based model using the Wannier functions is obtained for the system near the Fermi level. Some of the parameters to which the model is sensitive to are - (i) the energy window corresponding to which the model is to be formed and (ii) the projectors which contribute maximum within this energy window. The Wannier fitting obtained for YAuPb is shown in Fig. 3. The energy window used to obtain the fitting ranges from -4.5 to 4 eV around the Fermi level. The projections used in this process are $4d_{xy}$, $4d_{yz}$ and $4d_{xz}$ orbitals of Y, 6\textit{s} orbitals of Au and 6$p$ orbitals of Pb. It is seen from the figure that the Wannier bands (marked in red), obtained in this calculation, get nicely matched with that of the DFT bands (marked in black). This assures that the fitting obtained is reliable for the further calculation of the topological properties. The compound YAuPb is claimed to be topological semimetal in literature\cite{Vishwanath}. Motivated with this claim, some of the well-known phenomena observed in semimetals are extensively studied for the compound.

\begin{figure}[tbh]
  \begin{center}
    \includegraphics[width=0.75\linewidth, height=6.0cm]{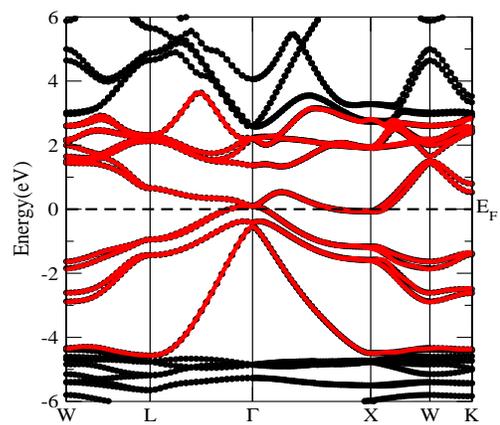} 
    \caption{The plot showing Wannier bands (red) getting nicely matched to DFT bands (black).}
    \label{fig:}
  \end{center}
\end{figure}

\begin{table*}
\caption{\label{tab:table1}%
\normalsize{The primitive coordinates of the node points along with their energy and chirality are presented below. The energy difference limits under which the nodes are obtained are also mentioned. The energy (E) of the node points are given with respect to Fermi level.
}}
\begin{ruledtabular}
\begin{tabular}{cccccccc}
\textrm{\textbf{S.No.}}&
\textrm{$\textbf{k}_{\textbf{1 }} \textbf{(\AA$^{\textbf{-1}}$)}$}&
\textrm{$\textbf{k}_{\textbf{2 }} \textbf{(\AA$^{\textbf{-1}}$)}$}&
\textrm{$\textbf{k}_{\textbf{3 }} \textbf{(\AA$^{\textbf{-1}}$)}$}&
\textrm{\textbf{gap (eV)}}&
\textrm{\textbf{E (eV)}}&
\textrm{\textbf{Chirality (C)}}&\\
\colrule
1  & -0.00059 &  0.00029 & -0.00002 & 0 & 0.10367 & -1   \\
2  &  0.00171 & -0.00022 & -0.00010 & 0 & 0.10356 &  1\\
3  & -0.00142 & -0.00285 & -0.00417 & 0 & 0.10394 & -1\\
4  & -0.00002 &  0.00006 & -0.00020 & 0 & 0.10359 &  1 \\
5  &  0.00003 & -0.00030 &  0.00005 & 0 & 0.10358 & -1\\
6  & -0.00729 & -0.00401 & -0.00579 & 0 & 0.10485 & -1 \\
7  &  0.00210 &  0.00029 &  0.00694 & 0 & 0.10457 &  1\\
8  &  0.00779 &  0.00261 &  0.00108 & 0 & 0.10349 & -1\\
9  & -0.00138 & -0.00258 & -0.00137 & 0 & 0.10381 &  1\\
10 & -0.00333 & -0.00327 & -0.00671 & 0 & 0.10433 &  1\\
\end{tabular}
\end{ruledtabular}
\end{table*}

\par One of the salient features of semimetals is the existence of characteristic node points. The node points between top of the valence bands and the bottom of the conduction bands are of great importance. This is because these points are associated with many topological properties, found commonly in semimetals \cite{Ashwin}. The compound, YAuPb, is tested for the presence of characteristic nodes. A total of 10 such nodes are obtained at $\sim$100 meV above the Fermi level. All these points are located in the vicinity of the \textit{$\Gamma$}-point. The details of the node points obtained are mentioned in Table I. The nodes in semimetals are associated with a specific chirality (C). The chirality of a node point situated at $\boldsymbol{k_0}$ is equal to $\frac{1}{2\pi}$ times the net Berry flux penetrating through any surface enclosing the point\cite{Ashwin}. This generally suggests that the positive and negative chiral nodes act as the source and sink of Berry flux, respectively. Furthermore, it is well known that the net chirality within the BZ must be zero. This implies that the nodes obtained must be in pairs of equal and opposite chiralities. The chiralities of the nodes obtained are also calculated and the result is mentioned in Table I. As can be seen from the table, a total of 5 pairs of nodes are obtained with equal and opposite chiralities. Hence, the result obtained is in terms of the fact that the net chirality across the BZ must be zero. It is already discussed that the nodes with a given non-zero chirality are the source or sink of Berry flux. Therefore, each node points acts as the monopoles of Berry curvature ($\boldsymbol\Omega$). In order to visualize it for a better understanding, the calculation of $\boldsymbol\Omega$ is performed.
\par $\boldsymbol\Omega$ is computed for a slab consisting of two characteristic nodes. For better realization of the effect, the nodes of opposite chiralities are chosen for present calculation. The calculation is performed using $7^{th}$ and $8^{th}$ nodes from Table I. The $\boldsymbol\Omega$ corresponding to these two nodes is shown in Fig. 4. The figure shows the projection of $\boldsymbol\Omega$ on $k_x$-$k_y$ plane of the BZ. The points \textquoteleft A' and \textquoteleft B' corresponds to the negative and positive chiral nodes, respectively. The direction of arrows at a given point represents the direction of $\boldsymbol\Omega$ vector at that point. It is seen from the figure that the $\boldsymbol\Omega$ diverges from negative chiral node, \textit{i.e.}, \textquoteleft A' and moves towards the positive chiral point. The curvature gets converged at the positive chiral node, \textit{i.e.}, \textquoteleft B'. Thus, the result obtained supports the fact that characteristic nodes act as the monopoles of the Berry curvature. Furthermore, it is well known that the topological materials usually have robust surface states\cite{Ashwin,Hasan}. As the compound YAuPb, shows topological behaviour, we are motivated to study its surface states.

\begin{figure}[tbh]
  \begin{center}
    \includegraphics[width=0.75\linewidth, height=5.5cm]{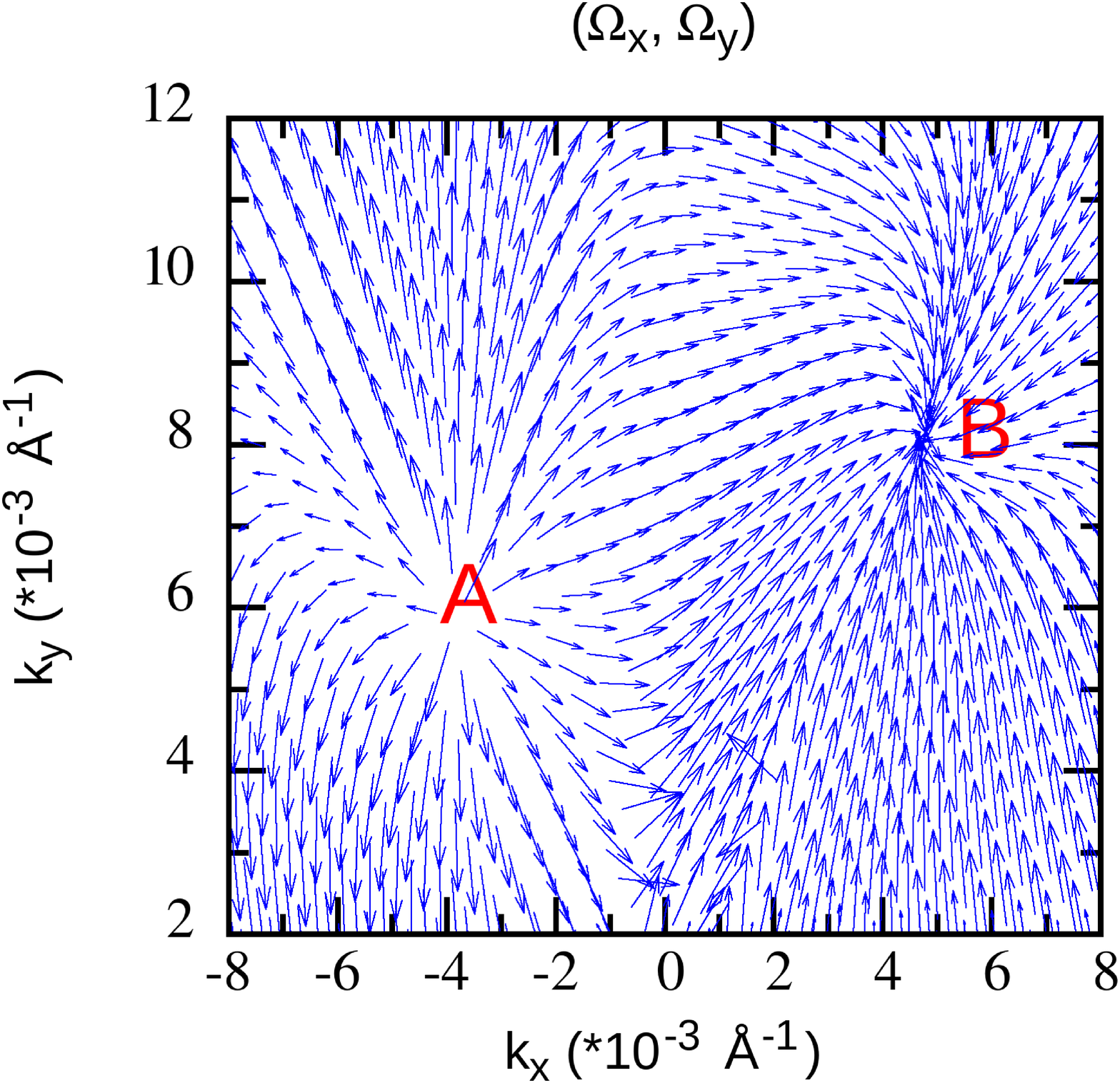}
    \caption{Berry Curvature corresponding to nodes $7^{th}$ and $8^{th}$ from Table I.}
    \label{fig:}
  \end{center}
\end{figure} 

\par The slab calculation of the surface states is also performed for the compound. For better visualization of these states, a dense k-mesh of 400 × 400 size in $k_1-k_2$ plane is used. The growth of the slabs are taken in the $k_3$ direction. The results obtained from this calculation are shown in Fig. 5, where the x- and y-axes are scaled in terms of the reciprocal lattice vector of the unit cell. Fig. 5 (a) shows the projection of node points from the bulk on the $k_1-k_2$ plane. The bluish and reddish regions respectively, indicate the absence and presence of node points in the perpendicular direction from the respective region of the plane, inside the bulk of the material. It is seen in the figure that instead of getting isolated points, a red circular region is obtained. This is due to the nodes being highly concentrated in a small region. Furthermore, Figs. 5 (b) and 5 (c) show the surface states for the slab systems corresponding to the top and bottom surfaces of the unit cell. In these figures, the bluish and reddish regions respectively, indicate the absence and presence of surface states in the $k_1-k_2$ plane. It is seen that the surface states for the slab system corresponding to top and bottom surfaces of the unit cell differs from one another. This seems to be because of different atomic arrangements near these surfaces. On close observation of these states, it is seen that they connect the projection points of opposite chiral nodes from the bulk on the $k_1-k_2$ plane. Such states are commonly referred to as Fermi arc. With the observation of surface states and other properties discussed above, the topological picture of the material becomes clear to a good extent. In the next section, the study of temperature dependent transport properties of YAuPb is carried out.

\begin{figure}
    \centering
    \subfigure[]
    {
        \includegraphics[width=0.75\linewidth,height=4.2cm]{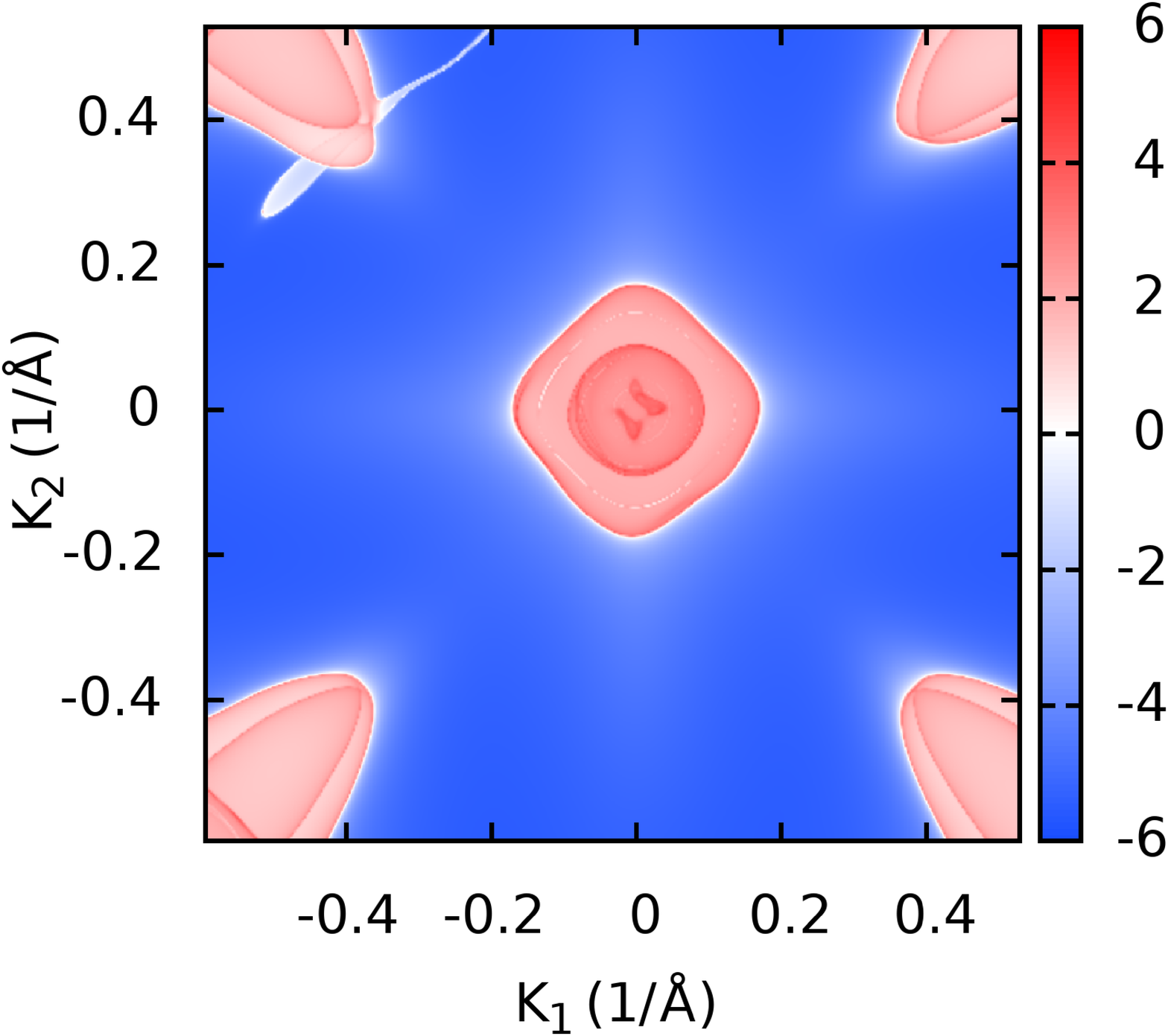}
    }
    \qquad
    \subfigure[]
    {
        \includegraphics[width=0.75\linewidth,height=4.2cm]{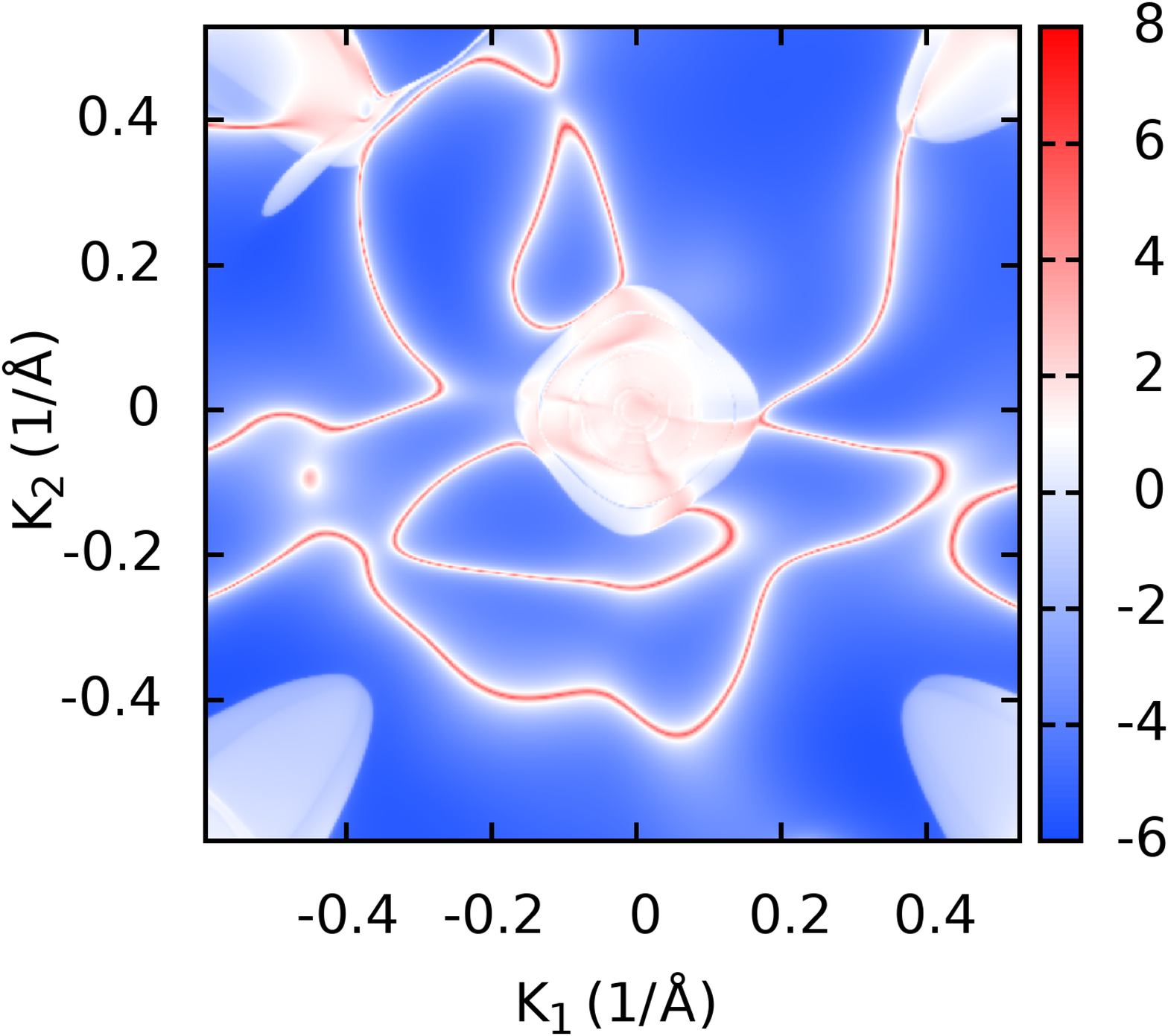}
    }
    \qquad
    \subfigure[]
    {
        \includegraphics[width=0.75\linewidth,height=4.2cm]{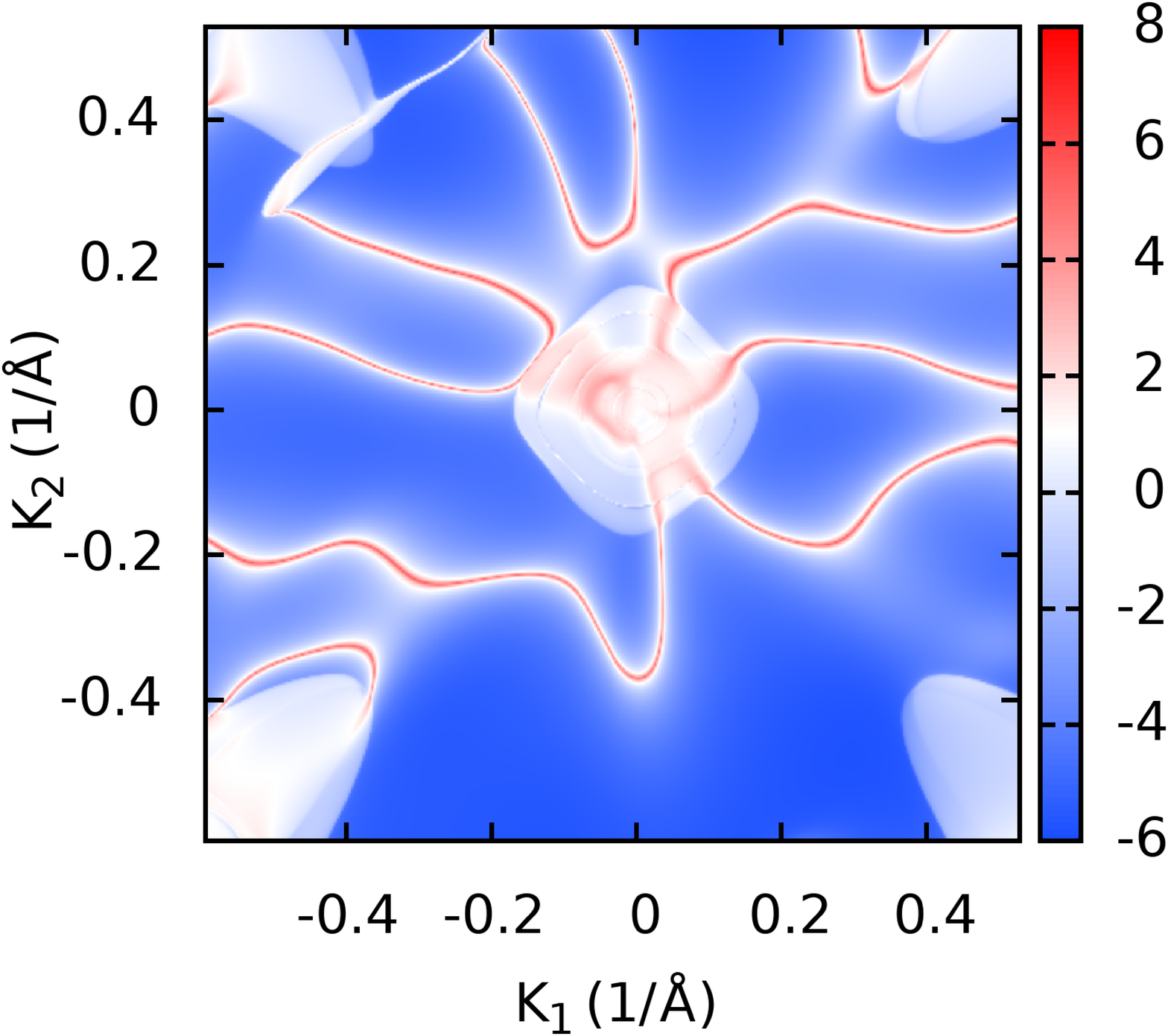}
    }
    \caption
    {
        \footnotesize (a) Nodes projection from bulk on the $k_1-k_2$ plane.
        (b) and (c) represents the surface states for the slab systems corresponding to the top and bottom surfaces of the unit cell respectively.
    }
    \label{fig:foobar}
\end{figure}

\subsection{Transport properties}

The electrical conductivity, $\boldsymbol\sigma$ is one of the important transport properties which helps in deciding the practical use of the compound. $\boldsymbol\sigma$ per unit relaxation time ($\boldsymbol {\sigma}/\tau$) is studied corresponding to $\mu$ at the node points ($\mu_N$) and the Fermi level ($\mu_F$). At this point, it is important to mention that the term \textquotedblleft $\mu_N$" in further discussions refers to the $\mu$ equals to $\sim$ 72.6 meV above the Fermi level. It is seen from Table 1 that the value of $\mu_N$ is close to the energy of all the nodes obtained. Thus, the calculations performed using $\mu = \mu_N$ is expected to be reliable for the explanation of the behaviour of any of the transport properties at the node points. The variation of $\boldsymbol {\sigma}/\tau$ with T, at both the $\mu$, is shown in Fig. 6 (a). At a given value of T, it is seen that the magnitude of $\boldsymbol {\sigma}/\tau$ is comparatively higher at the $\mu_N$ than $\mu_F$. It seems from the figure that the magnitude of $\boldsymbol {\sigma}/\tau$ remains constant with change in T. But in enlarged view, shown in Fig. 6 (b) and 6 (c), it is seen that there is small increase in its magnitude with the rise in T, at both the values of $\mu$. In Fig. 6 (b), representing the behaviour of the property at Fermi level, it is observed that the magnitude of $\boldsymbol {\sigma}/\tau$ is 
$\sim 1.36 \times 10^{19}$  $\Omega^{-1}m^{-1}s^{-1}$ at 50 K, which gets increased to $\sim 1.54 \times 10^{19} $  $ \Omega^{-1}m^{-1}s^{-1}$ at 300 K. On the other hand, in Fig. 6 (c) representing the behaviour at node points, it is seen that the magnitude of the property is 
$\sim 2.95 \times 10^{19}  $  $ \Omega^{-1}m^{-1}s^{-1}$ at 50 K, which is found to be raised to $\sim 3.08 \times 10^{19}$ $\Omega^{-1}m^{-1}s^{-1}$ at 300 K.

\begin{figure}[tbh]
  \begin{center}
    \includegraphics[width=2.8in]{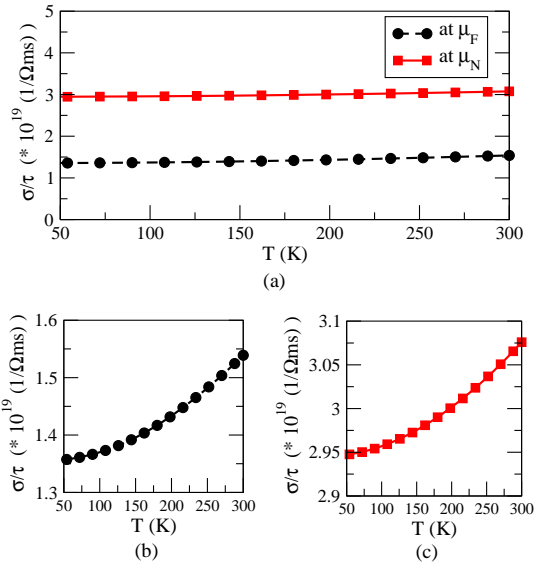}
    \caption{\footnotesize  (a) Electrical conductivity per unit relaxation time ($\boldsymbol {\sigma}/\tau$) versus Temperature (T) plot. The red (black) coloured curve represents the behaviour of $\boldsymbol {\sigma}/\tau$ at node points (Fermi level). (b) $\boldsymbol {\sigma}/\tau$ vs T at node points (c) $\boldsymbol {\sigma}/\tau$ vs T at Fermi level. }
    \label{fig:}
  \end{center}
\end{figure} 

To understand the behaviour obtained in Fig. 6, one needs to know the physical parameters upon which the value of $\boldsymbol\sigma$ in solids depends. It is known that $\boldsymbol\sigma=\boldsymbol\Sigma\boldsymbol{\sigma}_\textbf{n}$, where n denotes band indices. The expression for $\boldsymbol{\sigma}_\textbf{n}$ is given by \cite{Mermin},

\begin{equation}
  \boldsymbol{\sigma_n}=e^2 \int_{}^{} \frac{d\boldsymbol k}{4\pi^3} \tau_n(\varepsilon_n(\boldsymbol k)) \boldsymbol v_n(\boldsymbol k)\boldsymbol v_n(\boldsymbol k)\left[-\frac{\partial f}{\partial \varepsilon}\right]_{(\varepsilon_ = \varepsilon_n(\boldsymbol k) ) }
\end{equation}

where \textit{e} represents the electronic charge and $\tau_n(\varepsilon_n(\boldsymbol k))$ \& $\boldsymbol v_n(\boldsymbol k)$ are the relaxation time \& group velocity of the $n^{th}$ band, respectively. Also, the term $\frac{\partial f}{\partial \varepsilon}$ represents the partial derivative of Fermi-Dirac distribution function with respect to energy. Here $\varepsilon_n(\boldsymbol k)$ denotes the energy of a particular \textit{n}, \textbf{k} state.  The formula suggests that the magnitude of $\boldsymbol {\sigma_n}$ gets altered only with the change in the values of  -$\frac{\partial f}{\partial \varepsilon}$, $\tau_n(\varepsilon_n(\boldsymbol k))$, number of k-points at the given $\mu$ and $\boldsymbol v_n(\boldsymbol k)$. In the above relation, the integrand is significant for only those states, where the value of -$\frac{\partial f}{\partial \varepsilon}$ is non-zero. Hence it plays an important role in deciding the effective number of states contributing to the magnitude of $\boldsymbol{\sigma}_n$.

\begin{figure}[tbh]
  \begin{center}
    \includegraphics[width=0.80\linewidth, height=5.0cm]{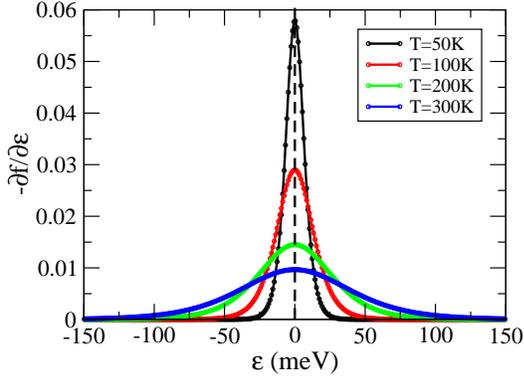}
    \caption{\footnotesize Variation of the function -$\frac{\partial f}{\partial \varepsilon}$ with energy ($\varepsilon$) at different temperatures.}
    \label{fig:}
  \end{center}
\end{figure} 

The variation of -$\frac{\partial f}{\partial \varepsilon}$ with energy ($\varepsilon$), at different temperatures, is shown in Fig. 7. The figure corresponds to $\mu$ = 0 meV which is equal to $\mu_F$. The variation will be similar at the $\mu_N$. It is seen from the figure that value of the function is non-zero only for the small energy region around the given $\mu$. This suggests that only the band-indices with k-points or the states having the energy within this region will contribute to the magnitude of $\boldsymbol {\sigma}$ at a particular temperature. The spread on the energy axis increases with the rise in temperature. Thus, the effective number of states contributing to the value of $\boldsymbol {\sigma}$ increases. In light of this, it is expected that at both the $\mu$, the value of $\boldsymbol {\sigma}$ will increase with the increasing temperature. Moreover, it is also seen that the constant-energy surface corresponding to $\mu_N$ is significantly higher than at the $\mu_F$. The respective plot corresponding to both the $\mu$ are shown in Fig. 8. The Fig. 8 (a) and 8 (b) indicates the constant energy surface of the first BZ at the $\mu_F$ and $\mu_N$, respectively. The dark blue-green region in these figures represents the regions of BZ with states having the energy equal to the respective values of $\mu$. The figure clearly indicates that a greater number of states fall within the effective energy interval at the $\mu_N$ than at the $\mu_F$. As a result, the magnitude of $\boldsymbol {\sigma}$ is supposed to be higher at the $\mu_N$ than at the $\mu_F$. It must be noted that the calculation of transport properties using BoltzTraP code is performed under constant relaxation time $(\tau_n(\varepsilon_n(\boldsymbol k)))$ approximation. Thus, the variation of $\boldsymbol \sigma$ and $\boldsymbol \sigma/\tau$ with the increase in T, must be similar in fashion. In light of this, the magnitude of $\boldsymbol \sigma/\tau$ at a given value of T must be higher at the $\mu_N$ than at the $\mu_F$. Also, at a given $\mu$, the magnitude of the $\boldsymbol \sigma/\tau$ must increase with the rise in T. Thus, the results of the calculation shown in Fig. 6, are consistent with the above discussion. Also, Gofryk \textit{et al.} performed the experimental study of the variation of resistivity ($\rho$) with T for the semimetallic half-Heusler alloys of class \textit{R}PdBi (\textit{R}=Er,Ho,Gd,Dy,Y,Nd)\cite{Gof}. In their work, a slow decrease in the value of $\rho$ in observed with the increasing T, for the temperature range of 50-300 K. For instance, for NdPdBi, the value of $\rho$ is $\sim 0.48 \hspace{1mm}(0.50)$ $m\Omega$V$ cm^{-1}$ at 50 (300) K. Corresponding value of $\boldsymbol\sigma$ will be $\sim 2.08 \hspace{1mm}(2.22)$ $\times 10^5$ $\Omega^{-1}$$m^{-1}$ at the given temperatures. This suggests that, within this temperature range, the value of $\boldsymbol {\sigma}$ for these materials will have slow increasing behaviour with the rising T. Thus, behaviour of $\boldsymbol {\sigma}$ in their works is consistent with the variation of $\boldsymbol\sigma/\tau$ observed for YAuPb, considering the $\tau$ of order $10^{-14}$.

\begin{figure}
    \centering
    \subfigure[]
    {
        \includegraphics[width=0.45\linewidth]{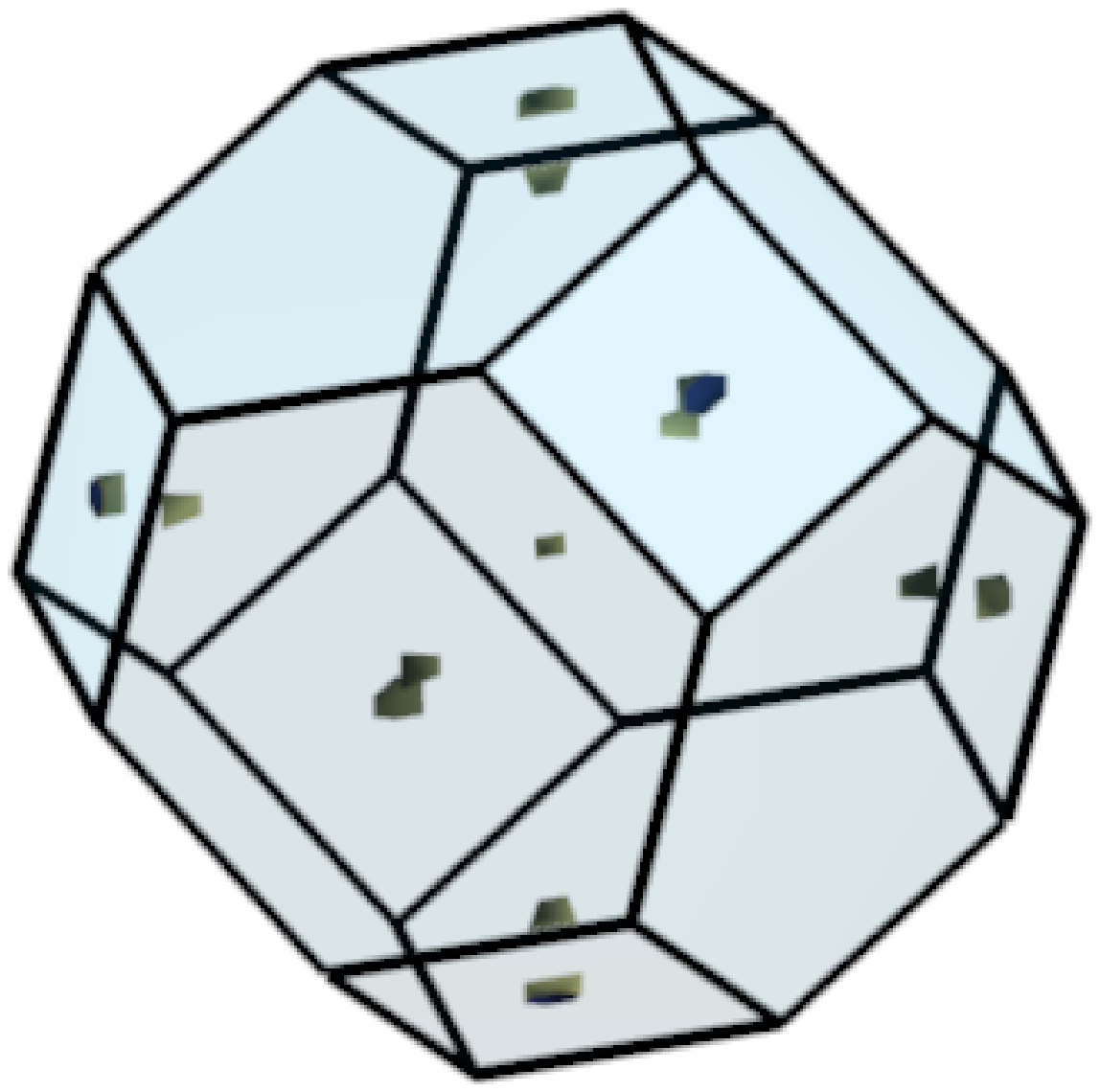}
    }
    \subfigure[]
    {
        \includegraphics[width=0.45\linewidth]{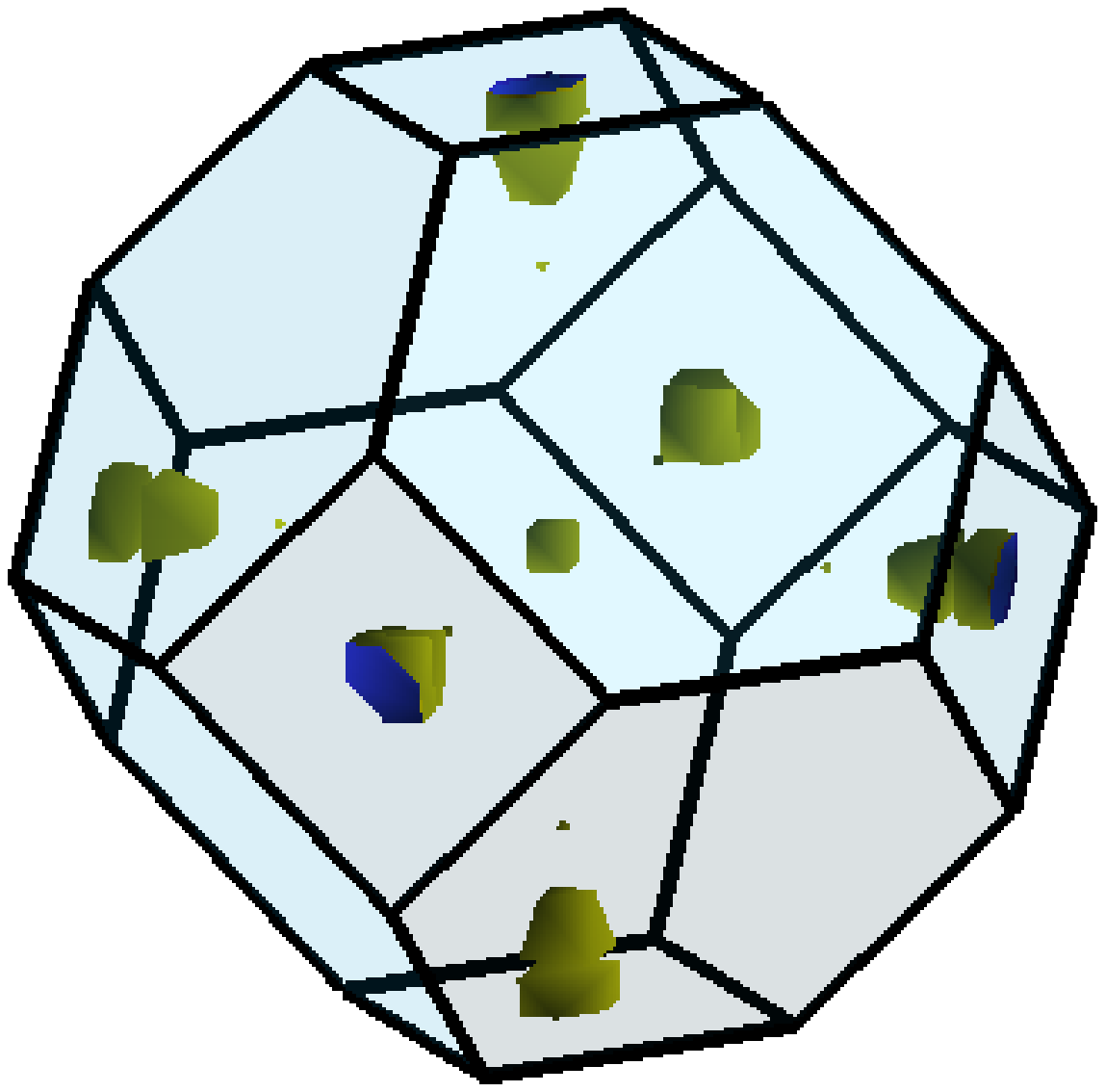}
    }
    \caption
    {
        (a) Constant energy surface at Fermi energy
        (b) Constant energy surface at Node point energy
    }
    \label{fig:foobar}
\end{figure}

\begin{figure}[tbh]
  \begin{center}
    \includegraphics[width=0.70\linewidth, height=4.5cm]{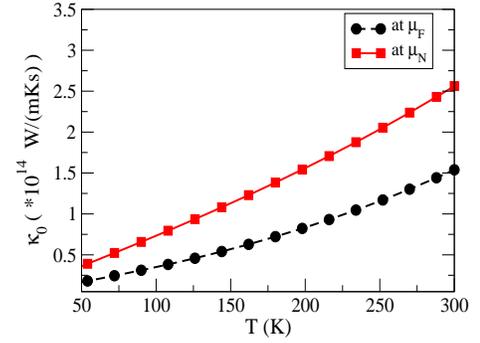}
    \caption{\footnotesize Electronic part of thermal conductivity per unit relaxation time ($\boldsymbol\kappa_0$) versus Temperature (T) plot. The red (black) coloured curve represents the behaviour of $\boldsymbol\kappa_0$ at node points (Fermi level).}
    \label{fig:}
  \end{center}
\end{figure}

Electronic part of thermal conductivity per unit relaxation time ($\boldsymbol\kappa_0$) is also studied for the compound at both the values of $\mu$. The variation of $\boldsymbol\kappa_0$ as a function of T, at both the $\mu$, is shown in Fig. 9 for the temperature range 50-300 K. It is seen from the figure that at a given $\mu$, the magnitude of $\boldsymbol\kappa_0$ increases in a linear fashion with the rise in T. At a particular T, the magnitude of $\boldsymbol\kappa_0$ is higher at the $\mu_N$ than at $\mu_F$ throughout the T range. The difference between its value at the two $\mu$ also increases with the rise in T. From the figure, it is observed that, at the $\mu_F$, the magnitude of $\boldsymbol\kappa_0$ is  $\sim 0.2 \times 10^{14}$ $ Wm^{-1}K^{-1}s^{-1}$ at 50 K, which gets increased to $\sim 1.5 \times 10^{14}$ $ Wm^{-1}K^{-1}s^{-1}$ at 300 K. On the other hand, at the $\mu_N$, its magnitude is  $\sim 0.4 \times 10^{14}  $  $ Wm^{-1}K^{-1}s^{-1}$ at 50 K, which gets changed to $\sim 2.6 \times 10^{14}$ $ Wm^{-1}K^{-1}s^{-1}$ at 300 K. In order to understand the behaviour obtained, it is important to know the physical parameters upon which the value of $\boldsymbol\kappa_0$ depends. In case of solids, the expression of $\boldsymbol\kappa_0$ is directly related to $\boldsymbol {\sigma}$ and is given by \cite{Mermin},

\begin{equation}
 \boldsymbol\kappa_0=\frac{\pi^2}{3}\left(\frac{k_B}{e}\right)^{2}T\left(\frac{\boldsymbol\sigma}{\tau}\right)
\end{equation}

where Boltzmann constant, $k_B$ and electronic charge, \textit{e} are physical constants. Also, the symbols T, $\boldsymbol\sigma$ and $\tau$ represent the temperature, electrical conductivity and relaxation time, respectively. The above relation suggests that the value of $\boldsymbol\kappa_0$ gets altered only with the change in values T and $\boldsymbol\sigma/\tau$.
It is already seen in Fig. 6 (b) and 6 (c) that, at a given $\mu$, the variation of $\boldsymbol\sigma/\tau$ with T is non-linear in fashion. Thus, it is expected from equation 2 that the magnitude of $\boldsymbol\kappa_0$ must not vary linearly with T, which is also evident from Fig. 9. Apart from this, it is seen in Fig. 6 (a) that, at a given T, the value of $\boldsymbol\sigma/\tau$ is higher at the $\mu_N$ than at the $\mu_F$. This behaviour seems to be the possible reason for the observation of higher magnitude of $\boldsymbol\kappa_0$ at $\mu_N$ than $\mu_F$. Moreover, the difference between the values of $\boldsymbol\kappa_0$, \textit{i.e.},($\Delta\boldsymbol\kappa_0$) at the two values of $\mu$, for a particular T is given by, $\Delta\boldsymbol\kappa_0=(\pi^2{k_B}^2/{3e^2})T\Delta({\boldsymbol\sigma}/{\tau})$. As can be seen from Fig. 6 (a) the difference between the magnitudes of $\boldsymbol\sigma/\tau$, (\textit{i.e.}, $\Delta(\boldsymbol\sigma/\tau)$) at the two $\mu$ remains almost constant to change in T. Thus, the expression of $\Delta\boldsymbol\kappa_0$ suggests that for the constant value of $\Delta(\boldsymbol\sigma/\tau)$, the value of $\Delta\boldsymbol\kappa_0$ should increase with the rise in T. The result obtained seems to be consistent with the above explanations. Furthermore, in the works of Singh \textit{et al.}, the study of temperature dependent variation of $\boldsymbol\kappa_0$ is carried out for YAuPb and LuAuPb\cite{Singh}. The behaviour and the values obtained in their results is similar to the results obtained for YAuPb in the present work. Along with $\boldsymbol\kappa_0$, the study of Seebeck effect (\textit{S}) also provides practical use of the compound.

The response of \textit{S} to the change in T is studied for the compound at both the $\mu$ and its variation is shown in Fig. 10 (a). The figure shows that the value of \textit{S} is negative at both the $\mu$. It is also observed that, at both the $\mu$, $|S|$ increases with the rise in T. For T$<$100 K, rate of increase in its magnitude is found to be approximately equal at both the $\mu$. But beyond 100 K, it is observed that as the T rises, rate of increase in the $|S|$ becomes comparatively higher at the $\mu_N$ than at the $\mu_F$. For understanding the behaviour obtained in the Fig. 10 (a), one needs to know the factors determining the value of \textit{S} in solid. The \textit{S} in solid is defined as \cite{Snyder},

\begin{equation}
  S=\frac{8\pi^2k_B^2}{3eh^2} T \left(\frac{\pi}{3n}\right)^{2/3}m^*
\end{equation}
where Boltzmann constant, $k_B$, electronic charge, \textit{e} and Planck's constant, h are physical constants. Also, the symbols \textit{n}, $m^*$ and \textit{T} represent the carrier concentration, effective mass and temperature, respectively. The above expression suggests that the value of \textit{S} gets altered only when there is a change in the values of \textit{n}, $m^*$ and \textit{T}. Under parabolic approximation, $m^*$ is defined as\cite{Mermin}, $m^*={\hbar}/{({\partial^2E}/{\partial k^2})}$ . The denominator in this expression represents the curvature of bands. Thus, the value of $m^*$ directly depends upon the band structure. It is implied from the above formula that the $m^*$ for the flat bands will be higher in comparison to that of curved bands. Also, the value of $m^*$ will be negative for electron pockets and positive for the hole pockets. Thus, the $m^*$ is useful in determining the type of carrier concentration in a material. It also decides the sign of \textit{S}. 

\begin{figure}
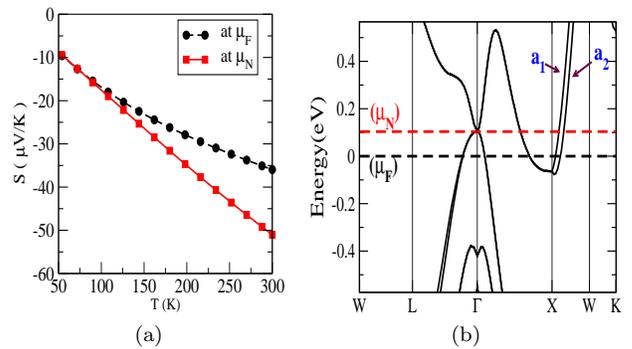

    \centering
    \subfigure[]
    {
        \includegraphics[width=0.42\linewidth, height=4.1cm]{fig10a.eps}
    }
    \subfigure[]
    {
        \includegraphics[width=0.48\linewidth, height=4.4cm]{fig10b.eps}
    }
    \caption
    {   (a) \footnotesize Seebeck coefficient (\textit{S}) versus Temperature (T) plot. The red (black) coloured curve       represents the behaviour of \textit{\textit{S}} at node points (Fermi level).
        (b) \footnotesize Dispersion curve showing the electron pocket in the vicinity of \textit{X}-point. The red (black) dashed-line indicates the energy corresponding to Node points (Fermi Level).
    }
    \label{fig:foobar}
\end{figure}

Fig. 10 (b) shows the dispersion curve in the region around the Fermi level. It is observed in the figure that there is an electron pocket in the vicinity of \textit{X}-point, with respect to both the $\mu$. The excitation of electrons from this region seems to be the possible reason for the negative values of \textit{S}. In \textit{X-W} direction (near \textit{X}-point), the figure shows that the band $a_2$ is flatter than band $a_1$. Thus, the $m^*$ of $a_2$ is expected to be higher than band $a_1$ in this region. Also, the band $a_1$ is at higher energy than band $a_2$ in $X$-$W$ direction. Therefore, from the concerned region, the electrons need lesser energy for getting excitation to move from band $a_1$ to a given higher $\mu$ than from band $a_2$. Hence, at lower values of T (\textit{i.e.} for T$<100$ K), the $m^*$ of only band $a_1$ contribute to the value of \textit{S} as the excitations from band $a_2$ is negligible. As the T is increased beyond 100 K, the excitations from band $a_2$ also becomes effective. As a result, the $m^*$ of both the bands contribute to the value of \textit{S}. Thus, at a given $\mu$, the value of $|S|$ is expected to increase with rising T. The result obtained is seen to be consistent with this. Furthermore, it is observed that the electron pocket corresponding to the $\mu_N$ is larger in size than that for the $\mu_F$. At lower values of T ($<100$ K), the electrons from the region close to the given $\mu$ get excited due to less energy gap. Thus, almost same number of electrons get excited to the $\mu_N$ and the $\mu_F$. As a result, for T$<100$ K, the values of \textit{S} is seen to be approximately equal for both the $\mu$. But as the T increases beyond 100 K, the electrons from the lower region of the electron pocket also get excited to the given $\mu$. The energy gap between the lower region of the electron pocket and the given $\mu$ is less for $\mu_F$ as compared to  $\mu_N$. Hence, a smaller number of electrons get excited to $\mu_N$ than at the $\mu_F$. Since, \textit{S} is inversely proportional to \textit{n}, it is expected that at higher values of T, $|S|$ will be more at $\mu_N$ than at the $\mu_F$. At T$>200$ K for the given T range, less number of electrons are expected to be available for excitation to $\mu_F$ from the electron pocket. This is because its size is comparatively smaller than the electron pocket corresponding to $\mu_N$ and thus the electrons available in it might be already excited as T is increased to 200 K. Thus, beyond 200 K within the given range of T, the rate of increase in the $|S|$  with the rise in T is supposed to be small at $\mu_F$ . On the other hand, due to larger size of electron pocket, this rate is expected to be comparatively higher at the $\mu_N$ for T beyond 200 K, within the given range of T. Hence, with the rise in T, the difference between the values of \textit{S} at both the $\mu$ is supposed to increase. The graph shown in Fig. 10 (a) seems to be in agreement with the above discussion. Also, Gofryk \textit{et al.} obtained the experimental values of \textit{S} for similar semimetallic half-Heusler alloys with the general formula \textit{R}PdBi (\textit{R}=Er,Ho,Gd,Dy,Y,Nd)\cite{Gof}. The range of $|S|$ obtained for these alloys is comparable to the values of $|S|$ computed for YAuPb. However, unlike YAuPb, the alloys \textit{R}PdBi have positive \textit{S} within the temperature range of 50-300 K. Thus, the material YAuPb has n-type behaviour whereas the alloys \textit{R}PdBi possess p-type behaviour in the range of 50-300 K.

\begin{figure}[tbh]
  \begin{center}
    \includegraphics[width=0.70\linewidth, height=4.5cm]{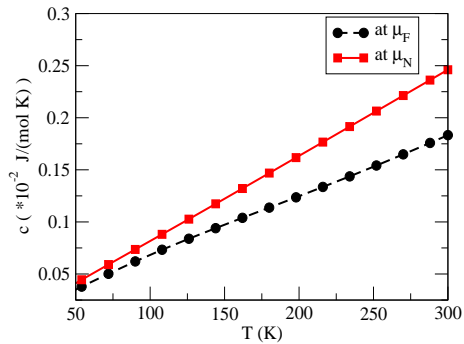}
    \caption{\footnotesize Electronic specific heat (\textit{c}) versus Temperature (T) plot. The red (black) coloured curve represents the behaviour of \textit{c} at node points (Fermi level)}
    \label{fig:}
  \end{center}
\end{figure}

The response of the electronic specific heat (\textit{c}) to T, at the $\mu_F$ and $\mu_N$ is studied for YAuPb, which are  depicted in Fig. 11. It is observed from the figure that the magnitude of \textit{c} increases with the rise in T, at both the values of $\mu$. It is also seen that at a particular value of T, the value of \textit{c} is higher at the $\mu_N$ than at the $\mu_F$. Moreover, the rate of increase in \textit{c} with the rise in T is also observed to be higher at the $\mu_N$ than at the $\mu_F$. To understand the behaviour obtained in the plot, it is important to study the factors upon which the property depends. The expression for \textit{c} in solids is given by\cite{Madsen},
\begin{equation}
 \textit{c}(T;\mu)=\int_{}^{}n(\varepsilon)(\varepsilon -\mu)\left[\frac{\partial f_\mu(T;\varepsilon)}{\partial T}\right] d\varepsilon
\end{equation}
where $\varepsilon$, $\mu$ and \textit{n($\varepsilon$)} denote the energy, chemical potential and density of states, respectively. The function $\frac{\partial f_\mu(T;\varepsilon)}{\partial T}$ represents the partial derivative of the Fermi-Dirac distribution function with respect to T at a particular $\mu$. The above expression suggests that, for a particular $\mu$, the integrand is significant in only those energy region where the value of $(\varepsilon -\mu)\left[\frac{\partial f_\mu(T;\varepsilon)}{\partial T}\right]$ is non-zero. This region is termed as the effective energy region, as only the states from this energy region will contribute to the value of \textit{c}. Thus, the term $(\varepsilon -\mu)\left[\frac{\partial f_\mu(T;\varepsilon)}{\partial T}\right]$ plays an important role in deciding the effective energy region. The variation of $(\varepsilon -\mu)\left[\frac{\partial f_\mu(T;\varepsilon)}{\partial T}\right]$  with $\varepsilon$ at the $\mu_F$ and $\mu_N$ are shown in Fig. 12 (a) and 12 (b), respectively. In addition to this, the density of states corresponding to the effective energy region at the $\mu_F$ and $\mu_N$ are shown in Fig. 12 (c) and 12 (d), respectively. It is observed from the Fig. 12 (a) and 12 (b) that the variation of the function is similar at both the $\mu$. Thus, the equation 4 suggests that difference in the values of \textit{c} at two $\mu$ is due to the difference in values of \textit{n($\varepsilon$)}. In addition to this, \textit{n($\varepsilon$)} in the effective energy region is higher for the $\mu_N$ than at the $\mu_F$. Therefore, the value of \textit{c} is expected to be always higher at the $\mu_N$ than at the $\mu_F$. Furthermore, the effective energy region is seen to be increasing with rise in T at both the $\mu$. As a result, a greater number of states contribute to the value of \textit{c} at higher T. It is evident from this behaviour that the value of \textit{c} must increase with the rise in T at a given $\mu$. Apart from this, it is seen in Fig. 12 (c) and 12 (d) that the magnitude of \textit{n($\varepsilon$)} is higher in the vicinity of $\mu_N$ than the $\mu_F$. Thus, the rate of increase in effective number of states with the rise in T is expected to be higher at $\mu_N$ than at the $\mu_F$. This suggests that the rate of increase in the value of \textit{c} with T must be higher for $\mu_N$ than for the $\mu_F$, which is reflected from the discussion of the Fig. 11. 

\begin{figure}[tbh]
  \begin{center}
    \includegraphics[width=1.00\linewidth, height=6.5cm]{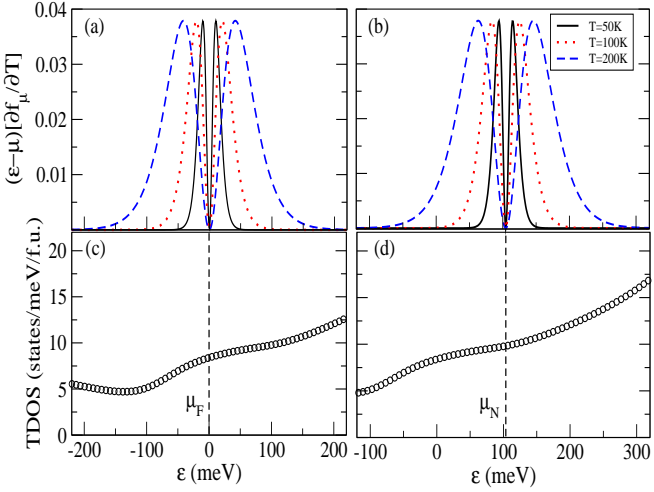}
    \caption{{\footnotesize The plot of $(\varepsilon -\mu)\left[\frac{\partial f_\mu(T;\varepsilon)}{\partial T}\right]$ versus $\varepsilon$ at $\mu_F$ and $\mu_N$ is shown in (a) and (b), respectively. Density of states in the respective energy region is shown in (c) and (d).}}
    \label{fig:}
  \end{center}
\end{figure}

\begin{figure}[tbh]
  \begin{center}
    \includegraphics[width=0.70\linewidth, height=4.5cm]{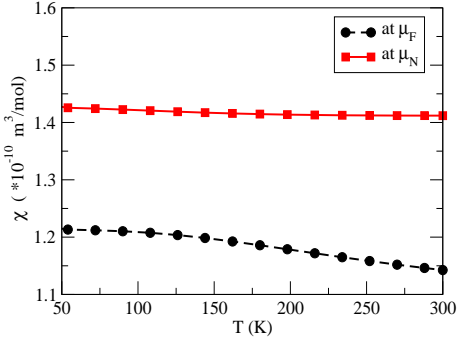}
    \caption{\footnotesize Pauli magnetic susceptibility ($\chi$) versus Temperature (T) plot. The red (black) coloured curve represents the behaviour of $\chi$ at node points (Fermi level).}
    \label{fig:}
  \end{center}
\end{figure}

The Pauli magnetic susceptibility ($\chi$) at different T is studied for the material with fixing the $\mu$ at $\mu_F$ and $\mu_N$, which are shown in Fig. 13. It is seen from the figure that the value of $\chi$ decreases with the rise in T, at both the $\mu$. It is also observed that the rate of decrease in $\chi$ with T is extremely small at the $\mu_N$. On the other hand, at the $\mu_F$, this rate is initially slow, but beyond T equals to 100 K, it gets increased.  The figure also depicts that, at a particular value of T, the value of $\chi$ is higher at the $\mu_N$ than at the $\mu_F$ in the given T range. In order to get an insight of the behaviour obtained, it is necessary to study $\chi$ in more details. The expression of $\chi$ is given by\cite{Madsen},
\begin{equation}
 \chi(T;\mu)=\mu_0\mu_B^2\int_{}^{}n(\varepsilon)\left[-\frac{\partial f_\mu(T;\varepsilon)}{\partial \varepsilon}\right] d\varepsilon
\end{equation}
where $\mu_0$, $\mu_B$ and \textit{$n(\varepsilon)$} stand for the permeability through free space, Bohr magneton and the density of states, respectively. The function $\frac{\partial f_\mu(T;\varepsilon)}{\partial \varepsilon}$ represents the partial derivative of the Fermi-Dirac distribution function with respect to $\varepsilon$ at a particular $\mu$. It is seen from the above expression that the value of $\chi$ gets altered only with the change in the values of \textit{$n(\varepsilon)$} and $\left[-\frac{\partial f_\mu(T;\varepsilon)}{\partial \varepsilon}\right]$, considering $\mu_0$ and $\mu_B$ as constant. Thus, the terms \textit{$n(\varepsilon)$} and $\left[-\frac{\partial f_\mu(T;\varepsilon)}{\partial \varepsilon}\right]$ play an important role in deciding the value of $\chi$. The variation of $\left[-\frac{\partial f_\mu(T;\varepsilon)}{\partial \varepsilon}\right]$ with $\varepsilon$ around the $\mu_F$ and $\mu_N$ will be similar to the plot shown in Fig. 7. In addition to this, the density of states corresponding to the effective energy region at both the values of $\mu$ are shown in Fig. 12 (c) and 12 (d), respectively. At the $\mu_F$, as observed from the Fig. 7, the peak value of the function $\left[-\frac{\partial f_\mu(T;\varepsilon)}{\partial \varepsilon}\right]$ decreases with the rise in T. Moreover, for $\varepsilon > \mu_F$ in the Fig. 12 (c), a small increase in \textit{n($\varepsilon$)} with the increase in \textit{$\varepsilon$} is seen. But for the $\varepsilon < \mu_F$, a sharp decrease in \textit{n($\varepsilon$)} is observed with decreasing \textit{$\varepsilon$}. Considering the overall effect of the change in \textit{n($\varepsilon$)}, a higher rate of decrease in the value of $\chi$ is expected (that at $\mu_N$, discussed below) with the rise in T. Unlike the case at $\mu_F$, it is observed in Fig. 12 (d) that \textit{$n(\varepsilon)$} increases with almost constant rate with the increasing $\varepsilon$, in the effective energy range. Since the plot of $\left[-\frac{\partial f_\mu(T;\varepsilon)}{\partial \varepsilon}\right]$ as a function of $\varepsilon$ is symmetric to the line $\varepsilon=\mu_N$, the effect on the value of $\chi$ due to increase in \textit{n($\varepsilon$)} at $\varepsilon= \mu_N +\delta \varepsilon $ is compensated by the decrease in $\chi$ due to the decrease in \textit{n($\varepsilon$)} at $\varepsilon= \mu_N -\delta \varepsilon $, where $\delta \varepsilon $ is a small energy interval. This suggests that the major contribution to the value of $\chi$ comes from the value of $\left[-\frac{\partial f_\mu(T;\varepsilon)}{\partial \varepsilon}\right]$ at $\varepsilon=\mu_N$, which is observed to be slowly decreasing with the rise in T. Thus, at $\mu_N$, a slow decrease in $\chi$ is expected with the increasing T. Also, in the works of Gofryk \textit{et al.}, the temperature dependent variation of $\chi$ is experimentally studied for the similar semimetallic half-Heusler alloy, YPdBi\cite{Gof}. The behaviour and the values obtained in their work is comparable to the results obtained for YAuPb.

\begin{figure}[tbh]
  \begin{center}
    \includegraphics[width=0.70\linewidth, height=4.5cm]{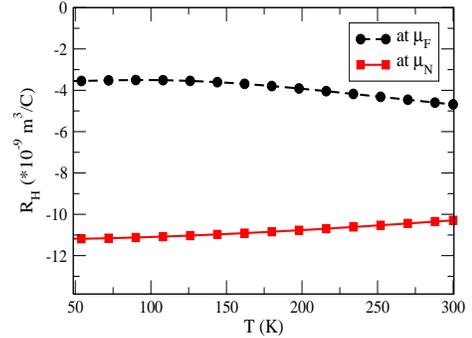}
    \caption{\footnotesize Hall coefficient ($R_H$) versus Temperature (T) plot. The red (black) coloured curve represents the behaviour of $R_H$ at node points (Fermi level)}
    \label{fig:}
  \end{center}
\end{figure}

The calculations of Hall coefficient ($R_H$) is also performed for the compound. Data points corresponding to change in $R_H$ with the rise in T are obtained corresponding to the $\mu_N$ and the $\mu_F$, which are shown in Fig. 14. It is seen in the figure that the values of $R_H$ are negative at both the $\mu$ in the given temperature range. Also, the value of $|R_H|$ corresponding to $\mu_N$ is higher than at the $\mu_F$. In addition to this, at the $\mu_F$ a small decrease in the value of $R_H$ is observed with the rise in T. On the other hand, at the $\mu_N$, a small increase in the value of $R_H$ is observed with the rise in T. Furthermore, it is needed to be mentioned that the disorder is normally present in the Heusler alloys. In present scenario, we have carried out all the calculations by considering the perfectly ordered YAuPb compound. The changes of magnetic and electronic structure properties due to the presence of various types of disordered in the Heusler alloy are explained by Feng \textit{et al.} in their work\cite{Feng}. Therefore in light of this, more detailed studies should be carried out for seeing whether the topological and thermoelectric properties of YAuPb will also show drastic change or not with the presence of different types of disorder in the material.

\section{Conclusions} 
 
\par In the present work, the study of electronic properties along with the detailed analysis of the topological behaviours and transport properties of YAuPb have been performed. In the analysis of the DFT bands, the inversion of band characters is observed near the Fermi level. The orbitals responsible for the band-inversion are found to be as \textquoteleft 5\textit{p}' and  \textquoteleft 6\textit{s}' orbitals of Pb and Au, respectively. In case of topological properties, 5 pairs of characteristic nodes having equal and opposite chiralities are obtained, at $\sim$ 100 meV above the Fermi level. The calculations also showed that these nodes act as monopoles of Berry curvature. Besides this, several surface states corresponding to the first Brillouin zone are calculated. Based on these results, the material can be categorised as non-trivial topological semimetal. Apart from this, the computed value of the electrical conductivity per unit relaxation time ($\boldsymbol {\sigma}/\tau$) corresponding to the Fermi level ($\mu_F$) is found to be $\sim 1.36 \hspace{1mm}(1.54)\times 10^{19}$  $\Omega^{-1}m^{-1}s^{-1}$ at 50 (300) K . Also, the value of electronic part of thermal conductivity per unit relaxation time ($\boldsymbol\kappa_0$) corresponding to $\mu_F$ is found to be  $\sim 0.2 \hspace{1mm}(1.5) \times 10^{14}$ $ Wm^{-1}K^{-1}s^{-1}$ at 50 (300) K. These values indicate the conducting nature of the material to both heat and electricity. Furthermore, the computed value of \textit{S} corresponding to $\mu_F$ is found to be $\sim -9.07 \hspace{1mm}(-35.95)$ $\mu$V$ K^{-1}$ at 50 (300) K, which indicates the n-type behaviour of the material. At the $\mu_F$, the value of electronic specific heat (Pauli magnetic susceptibility) is obtained to be $\sim 0.03 \hspace{1mm}(0.18) \times 10^{-2}$ $ Jmol^{-1}K^{-1}$ ($\sim 1.21 \hspace{1mm}(1.14) \times 10^{-10}$ $ m^{3}mol^{-1}$) at 50 (300) K. Thus, our work suggests that the material YAuPb is semimetallic in nature, having the potential to be employed in the transmission of both heat and electricity, and can also be used as n-type material in the low temperature range (50-300 K). Furthermore, if experimentally we need to raise the Fermi level to $\mu_N$, an electron doping of concentration equals to $1.16\times 10^{21}$ $electrons/cm^3$ is required.

\end{document}